\documentclass[
  aip,
  pop,
  amsmath,amssymb,
  reprint,
]{revtex4-1}
\usepackage{graphicx,color}
\usepackage{amsmath}
\usepackage{natbib}
\usepackage{epsfig}
\begin{document}

\title{Note: Sound velocity of a soft sphere model near the fluid-solid phase transition}

\author{Sergey A. Khrapak}
\affiliation{Aix-Marseille-Universit\'{e}, CNRS, Laboratoire PIIM, UMR 7345, 13397 Marseille cedex 20, France;
Forschungsgruppe Komplexe Plasmen, Deutsches Zentrum f\"{u}r Luft- und Raumfahrt (DLR),
Oberpfaffenhofen, Germany;
Joint Institute for High Temperatures, Russian Academy of Sciences, Moscow, Russia}

\begin{abstract}
The quasilocalized charge approximation is applied to estimate the sound velocity of simple soft sphere fluid with the repulsive inverse-power-law interaction. The obtained results are discussed in the context of the sound velocity of the hard-sphere system and of liquid metals at the melting temperature.    
\end{abstract}

\maketitle

It is well recognized that the ratio of the sound to thermal velocity of many liquid metals and metalloids has about the same value at the melting temperature,~\cite{IidaBook,BlairsPCL_2007,RosenfeldJPCM_1999}
\begin{equation}\label{1}
c_{\rm s}/v_{T}|_{T=T_{\rm m}}=\chi\sim 10. 
\end{equation}
Here $c_{\rm s}$ is the sound velocity, $v_{T}=\sqrt{T/m}$ is the thermal velocity, $T$ is the temperature in energy units, $m$ is the atomic mass, and $T_{\rm m}$ is the melting temperature. The values of $\chi$ for 41 liquid metals and metalloids tabulated in Ref.~\onlinecite{BlairsPCL_2007} agree well with Eq.~(\ref{1}), with extreme deviations $\chi_{\rm max}\simeq 15.3$ for Hg and $\chi_{\rm min}\simeq 4.1$ for Te.

Rosenfeld pointed out that this ``quasi-universal'' property of liquid metals is also shared by the hard-sphere (HS) model. In particular, starting from the thermodynamic definition of the fluid sound velocity,~\cite{LL_Fluids} $c_{\rm s}=\sqrt{(\partial P/\partial \rho)_S}$, where $P$ and $\rho$ are the pressure and mass density ($\rho=nm$), and the derivative is taken at constant entropy $S$, he derived the expression
\begin{equation}
c_{\rm s}/v_T=\left[p(\eta)+\eta dp(\eta)/d\eta+\tfrac{2}{3}p(\eta)^2\right]^{1/2}.
\end{equation}
Here $p(\eta)$ is the reduced pressure and $\eta=(\pi/6)n\sigma^3$ is the HS packing fraction, $\sigma$ being the hard sphere diameter. With the Carnahan-Starling equation of state (EoS) he then obtained $\chi\simeq 12.6$ at  $\eta_{\rm F}=0.494$ corresponding to the fluid-solid transition. This value of $\chi$ is rather insensitive to the concrete form of the EoS. For example, using the Percus-Yevick compressibility route EoS we get $\chi\simeq 13.4$, and the virial expansion with the first ten virial coefficients retained~\cite{SantosPRL_2012} yields $\chi\simeq 12.3$.

The purpose of this Note is to demonstrate that the property displayed by Eq.~(\ref{1}) is also exhibited by the most simple soft sphere model. The soft sphere model adopted here corresponds to the system of point-like particles interacting via the repulsive inverse-power-law (IPL) potential, $V(r)=\epsilon(a/r)^{\alpha}$, where $\epsilon$ is the energy scale and $a$ is the length scale, which is taken equal to the Wigner-Seitz radius, $a=(4\pi n/3)^{-1/3}$. We consider the regime $\alpha>3$, so that the neutralizing background is not necessary. Previously, the soft sphere model with the adjusted attractive interaction was employed to calculate the acoustic velocity using the standard thermodynamic definition.~\cite{ShanerJCP_1988} Here a different approximate approach is used, which results in particularly   simple relation between the sound velocity and thermodynamic properties.  

To estimate the sound velocity in the strongly interacting soft sphere system we use the quasilocalized charge approximation (QLCA).~\cite{GoldenPoP2000} In the last two decades this approach has been successively applied to several strongly coupled systems, mainly in the plasma-related context. This includes one-component-plasma (OCP) in both three (3D) and two (2D) dimensions,~\cite{GoldenPoP2000} complex (dusty) plasmas with screened Coulomb (Yukawa) interactions in 3D and 2D,~\cite{RosenbergPRE1997,KalmanPRL2000,
KalmanPRL2004,DonkoJPCM2008} complex plasmas with Lennard-Jones-like interactions in 3D,~\cite{RosenbergCPP2015} and 2D systems with dipole-like interactions.~\cite{GoldenPRB_2008,GoldenPRE2010} Good agreement between the dispersion relations obtained via QLCA and molecular dynamics simulations has been documented.~\cite{KalmanPRL2000,KalmanPRL2004,DonkoJPCM2008,OhtaPRL2000,KhrapakPoP2016}   
Recently, it has also been demonstrated that for weakly screened Yukawa fluids the longitudinal sound velocities evaluated using the QLCA are in good agreement with those calculated using the thermodynamic route (thermodynamic values being normally by several percent lower).~\cite{KhrapakPoP2016,KhrapakPRE2015_Sound,Semenov_2DSound,
KhrapakPPCF2015}

\begin{figure}
\includegraphics[width=7.2cm]{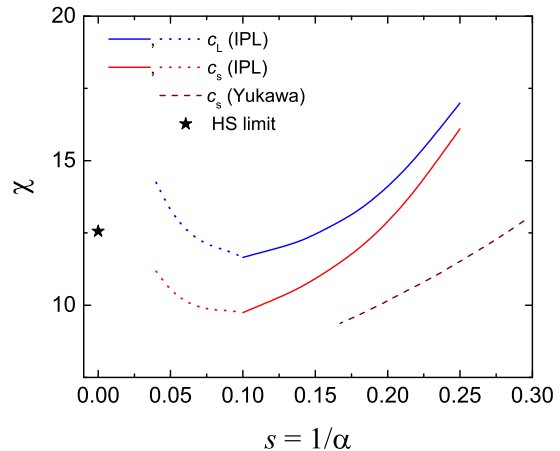}
\caption{The coefficient $\chi=c_{\rm L,s}/v_T$ at the melting temperature as a function of the potential softness parameter $s=1/\alpha$, with $\alpha=1+\kappa$ for the Yukawa interaction. The blue (red) curve corresponds to the elastic, $c_{\rm L}$, (thermodynamic, $c_{\rm s}$) sound velocity of the IPL system. The dashed curve corresponds to the thermodynamic sound velocity of Yukawa systems at melting. The symbol at $s=0$ is the HS result by Rosenfeld.~\cite{RosenfeldJPCM_1999}}
\label{Fig1}
\end{figure}

The generic QLCA expression for the longitudinal mode dispersion in a one-component system of (non-charged) particles interacting via an isotropic pairwise potential $V(r)$ coincides with that of Hubbard and Beeby~\cite{Hubbard1969} and reads 
\begin{equation}
\omega_{\rm L}^2=\frac{n}{m}\int\frac{\partial^2 V(r)}{\partial z^2} g(r) \left[1-\cos(kz)\right]d{\bf r},
\end{equation} 
where $g(r)$ is the equilibrium radial distribution function, $k$ is the wave number, and $z=r\cos\theta$ is the direction of the propagation of the longitudinal mode. Substituting the IPL potential we get in the long-wavelength limit, $k\rightarrow 0$, the acoustic dispersion with the sound velocity
\begin{equation}\label{cs1}
c_{\rm L}^2=\frac{1}{30}\omega_0^2a^2\alpha(3\alpha+1)\int_0^{\infty}x^{2-\alpha}g(x)dx,
\end{equation} 
where $\omega_0=\sqrt{4\pi n \epsilon a/m}$ is the nominal frequency and $x=r/a$. The integral in Eq.~(\ref{cs1}) is related to the reduced excess pressure, $p_{\rm ex}=P/nT-1$, of the IPL system: 
\begin{equation}\label{pex}
p_{\rm ex}=\frac{1}{6}\frac{\omega_0^2a^2\alpha}{v_T^2}\int_0^{\infty}x^{2-\alpha}g(x)dx.
\end{equation} 
Combining equations (\ref{cs1}) and (\ref{pex}) we immediately obtain a simple relation between the sound velocity and excess pressure of the IPL fluid
\begin{equation}\label{cL}
c_{\rm L}=v_T\sqrt{p_{\rm ex}(3\alpha+1)/5}.
\end{equation}
The longitudinal sound velocity obtained in this way can be referred to as the elastic sound velocity. Another quantity, perhaps more appropriate for the comparison with the HS results by Rosenfeld, is the thermodynamic  (or instantaneous~\cite{Schofield1966}) sound velocity defined by $c_{\rm s}^2=c_{\rm L}^2-\tfrac{4}{3}c_{\rm T}^2$, where $c_{\rm T}$ is the transverse sound velocity.~\cite{LL} The transverse sound velocity can also be evaluated using the QLCA, but this is not required here in view of the general relationship~\cite{Schofield1966,KhrapakRel} $c_{\rm L}^2-3c_{\rm T}^2=2p_{\rm ex} v_T^2$, yielding 
\begin{equation}\label{cS}
c_{\rm s}=v_T\sqrt{p_{\rm ex}(\alpha+3)/3}.
\end{equation}
Equations (\ref{cL}) and (\ref{cS}) apply to strongly coupled fluids, because QLCA is the theory for a strongly coupled state. In particular, it should apply to the IPL fluid near the fluid-solid phase transition. We have, therefore, used the data for the pressure and fluid density at the fluid-solid coexistence of the IPL system tabulated in Ref.~\onlinecite{Agrawal_1995} and estimated the coefficient $\chi$ from Eqs.~(\ref{cL}) and (\ref{cS}). The results are shown in Fig.~\ref{Fig1}. There is a wide range of potential softness where the ratio $\chi$ is practically constant:
For $6<\alpha<20$ we have $\chi\simeq 12.5\pm 1$ (using $c_{\rm L}$) and $\chi\simeq 10.5\pm 1$ (using $c_{\rm s}$). This is rather close to the value predicted by the HS model (shown by the symbol at $s=0$). As the interaction softens, $c_{\rm L}$ and $c_{\rm s}$ approach each other, as expected. Also shown by the dashed line is the thermodynamic sound velocity $c_{\rm s}$ of the repulsive Yukawa system [$V(x)=\epsilon e^{-\kappa x}/x$, where $\kappa$ is the screening parameter] at melting, obtained using the fluid approach of Ref.~\onlinecite{KhrapakPRE2015_Sound}. To produce this curve a simplest (but perhaps not the best) relation between the softness parameters of the IPL and Yukawa systems, $\alpha=1+\kappa$, has been used.~\cite{KhrapakPRL2009} The results are only shown for the regime, where reliable data on the thermodynamic properties are available ($\kappa\lesssim 5$). Here the values $c_{\rm L}$ and $c_{\rm s}$ are relatively close to each other and fall close to the range of interest.       
As $\kappa$ decreases, the ratio $\chi$ increases monotonously  and diverges in the OCP limit ($\kappa=0$, $\alpha=1$), where the dispersion becomes~\cite{GoldenPoP2000} $\omega_{\rm L}\simeq \omega_0$. 
 
An important point raised by this consideration is the applicability limit of the QLCA from the side of HS-like interactions. Namely, Eqs.~(\ref{cL}) and (\ref{cS}) imply $c_{\rm L,s}\propto \sqrt{\alpha p_{\rm ex}}$ as $\alpha\rightarrow \infty$. However, the excess pressure remains finite in the HS limit~\cite{Agrawal_1995} as does the sound velocity. This contradiction is a strong indication that QLCA loses its applicability as the softness of the interaction potential decreases. This is why the dotted curves have been used to depict the QLCA result at $s\lesssim 0.1$ in Fig.~\ref{Fig1}. More accurate location of the applicability limit of the QLCA will be a subject of future work.  

Finally, it should be pointed out that neither the simplistic soft sphere model nor the hard sphere model can provide detailed agreement with the thermodynamics and related properties of much more complex real systems (like e.g. liquid metals). Nevertheless, they can indicate the origin of some ``quasi-universal'' properties of such systems, like for instance the sound velocity near the fluid-solid transition discussed here.         

This work was supported by the A*MIDEX grant (Nr.~ANR-11-IDEX-0001-02) funded by the French Government ``Investissements d'Avenir'' program.

\bibliographystyle{aipnum4-1}
\bibliography{References_KhrapakSound}

\end{document}